\documentclass[a4paper,twocolumn]{article}

\AtBeginDocument{\addtolength{\columnsep}{8pt}}
\AtBeginDocument{\addtolength{\textwidth}{10pt}}
\usepackage{amsmath,amsfonts,amssymb}
\usepackage{amsthm}
\usepackage{enumerate}
\usepackage{bbm}
\usepackage{graphicx}
\usepackage{lscape,epsf}
\usepackage[vcentermath]{youngtab}
\usepackage{rotating}
\usepackage{cite}
\usepackage{comment}

\def\ds{\displaystyle}
\def\bea{\begin{array}{c}}
\def\ea{\end{array}}
\def\be{\begin{equation}\bea\ds}
\def\ee{\ea\end{equation}}
\def\bee{\begin{equation}\begin{array}{rcl}\ds}
\def\eee{\end{array}\end{equation}}

\def\hbar{g_s}

\def\chs[#1]{S_{#1}}
\def\L{\mathcal{L}}
\def\nav[#1]{\langle\!\langle{#1}\rangle\!\rangle}
\def\wJ[#1]{\widehat{J_{[#1]}}}

\title{Matrix integral expansion of colored Jones polynomials for figure-eight knot}

\author{A.\,Alexandrov$^{*+}$ and
D.\,Melnikov$^{+\dagger}$}

\date{}

\begin{document}

\twocolumn[
  \begin{@twocolumnfalse}
  \hfill{ITEP-TH-38/14}
    \maketitle
    \vspace{-0.5cm}
    \begin{center}
    {\itshape $^*$Mathematics Institute, University of Freiburg, \\ Eckerstra{\ss}e 1, D-79104 Freiburg, Germany\\
    \vspace{0.2cm}
    $^+$Institute for Theoretical and Experimental Physics, \\ B.~Cheremushkinskaya 25, Moscow 117218, Russia\\
    \vspace{0.2cm}
    $^\dagger$International Institute of Physics, UFRN, \\ Av. Odilon Gomes de Lima 1722, Capim Macio, Natal-RN  59078-400, Brazil}
    \end{center}
    \vspace{0.2cm}
    \begin{abstract}
In this note we examine a possible extension of the matrix integral representation of knot invariants beyond the class
of torus knots. In particular, we study a representation of the $SU(2)$ quantum Racah coefficients by double matrix integrals. We
find that the Racah coefficients are mapped to expansion coefficients in some basis of double integrals. The transformed
coefficients have a number of interesting algebraic properties.
\end{abstract}
\vspace*{0.5cm}
  \end{@twocolumnfalse}
  ]

Classification of knots is a central problem of knot theory. One way to solve it is to derive most general formulae for the
topological invariants, which would distinguish any pair of knots. For torus knots a general formula, computing the corresponding
HOMFLY polynomials, was derived by Rosso and Jones~\cite{Rosso:1993vn}. Since the work of Rosso and Jones a number of new general
formulae was derived (\emph{e.g.}~\cite{GeneralFormulae}-\cite{Nawata:2013qpa} and references therein), which yield (colored)
Jones or HOMFLY polynomials of certain knot series.

One possible way to extend the known results would be to construct a matrix integral representation of the knot invariants. The
relevant matrix integral for the Rosso-Jones formula was derived in~\cite{matrixmodel,Brini:2011wi}. Per se, the restriction to
the Rosso-Jones formula seems to be solely a technical issue and the authors are unaware of any conceptual obstacles to
generalize this result to a more general class of knots or links. Indeed, the existence of the matrix model formulation is
advocated by the relation of the knot invariants to the generic Hurwitz $\tau$-functions, as observed
in~\cite{Alexandrov:2014zfa}.

In a recent work~\cite{Alexandrov:2014nla} a step towards a generalization of the matrix model beyond the torus knots was taken.
For a series of twisted knots it was demonstrated that it is natural to consider matrix integrals with the integration measure
made of the Laplace evolution of the corresponding Jones polynomial. Unfortunately, this approach requires some unknown
ingredient, and the naive matrix integrals give a discrepancy in the perturbative expansion of HOMFLY polynomials starting from
$g_s^5$ terms.

The topological quantum field theory (TQFT) approach~\cite{Witten:1988hf} allows one to write generic formulae for knots
invariants. However, in this case the final result depends on the quantum extensions of the Racah coefficients (Wigner
$6j$-symbols). These coefficients are known in the case of $SU(2)$~\cite{Kirillov:1991ec,Itoyama:2012re}, and for symmetric and
antisymmetric representations of $SU(N)$~\cite{Itoyama:2012re,Nawata:2013ppa}, but remain largely unknown beyond those
results~(see~\cite{Zodinmawia:2011oya,Mironov:2011aa,Anokhina:2012ae,Anokhina,Gu:2014nva} for some exceptions), especially for
the cases of non-trivial multiplicity.

Based on the TQFT approach, in this work we study a matrix integral representation of the quantum Racah coefficients, which
naturally arises in the analysis of the invariants of the figure-eight ($4_1$) knot. The main idea behind our approach is to
consider these invariants as deformations of those of the product of a Hopf link and its mirror image. This will lead to an
expansion of the invariants in terms of a basis of double matrix integrals. We found that the coefficients of the expansion can
be thought as of a transform of the original Racah coefficients. To explain this idea we will rely on an explicit realization of
Witten's proposal developed in~\cite{Kaul:1991np} and subsequent papers, \emph{e.g.} see~\cite{Ramadevi,
Zodinmawia:2011oya,Nawata:2013qpa}. A nice refined review of the method was recently given in~\cite{Gu:2014nva}.

\begin{figure}
 \centering

\includegraphics[width=0.5\linewidth]{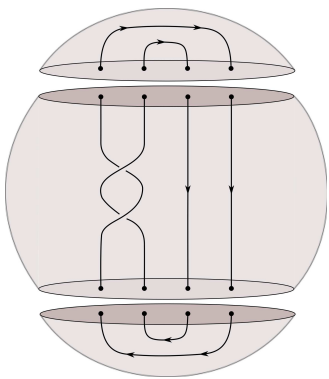}

 \caption{The topological invariant of the Hopf link $2_1^2$ can be obtained by sandwiching the braid group element $b_1^{-2}$
 between the initial and final state.}
 \label{fig_link}
\end{figure}

In the TQFT approach the invariant (HOMFLY polynomial of variables $q$ and $A\equiv q^N$) of the Hopf link $2_1^2$ is given by
the formula
\begin{multline}
\label{212Rmatrix}
H_{R_1,R_2}(2_1^2)\propto\sum\limits_{R_s\in R_1\otimes R_2}  \langle\, \psi_s\,|\, \dim_q R_s \,b_1^{-2}\,|\,\psi_s\,\rangle
\\ \propto \sum\limits_{R_s\in R_1\otimes R_2}
\dim_q R_s\, q^{2C_2(R_s)}  \,,
\end{multline}
where $R_1$ and $R_2$ are representations associated to the components of the link, $\dim_q R$ and $C_2(R)$ are the quantum
dimension and quadratic Casimir of the representation $R$. Figure~\ref{fig_link} illustrates the above formula in the TQFT
approach.

\begin{figure}[t]

\begin{minipage}{0.40\linewidth}
\centering
 \includegraphics[width=\linewidth]{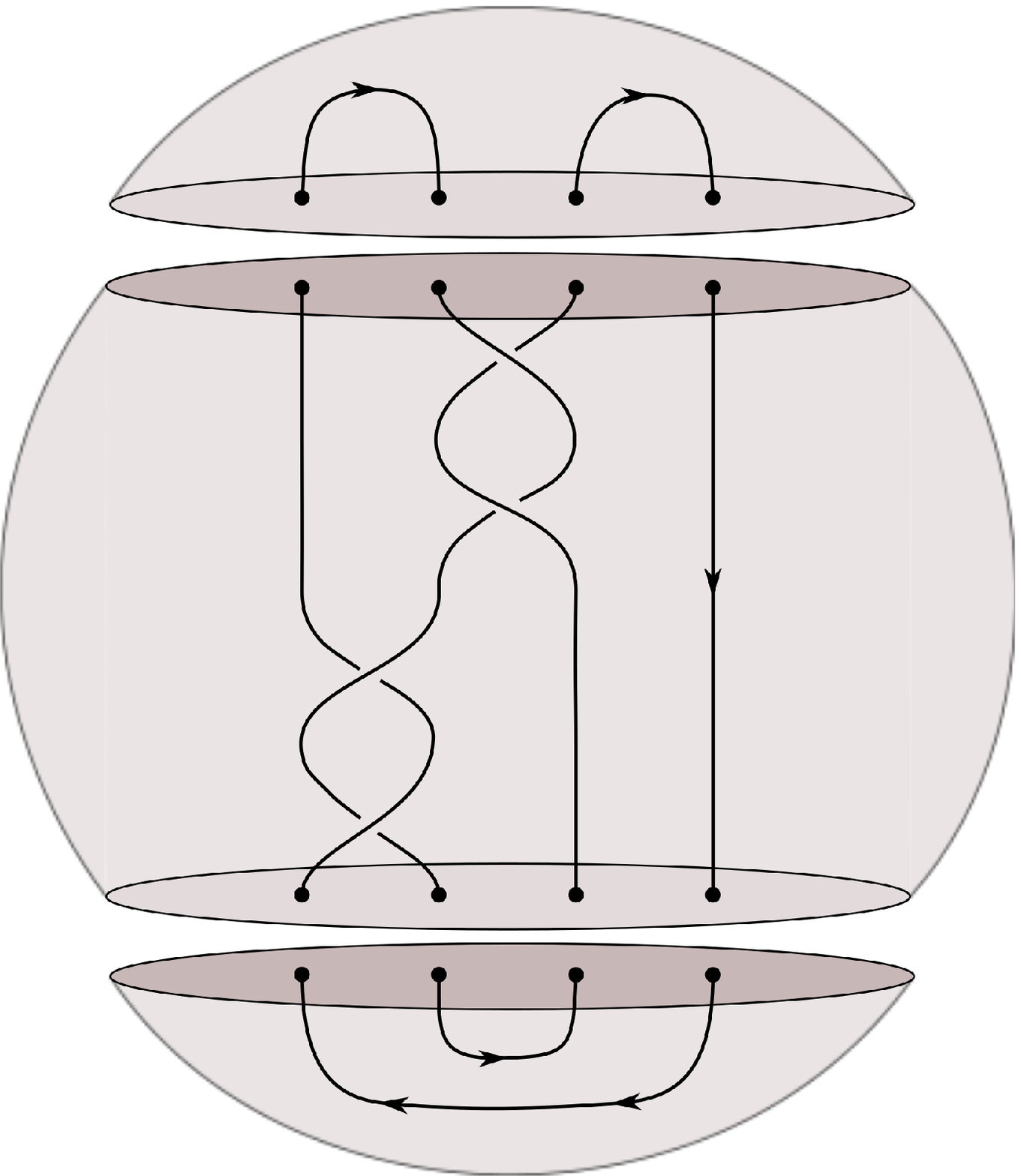}
\end{minipage}
\hfill{
\begin{minipage}{0.55\linewidth}
\centering
 \includegraphics[width=\linewidth]{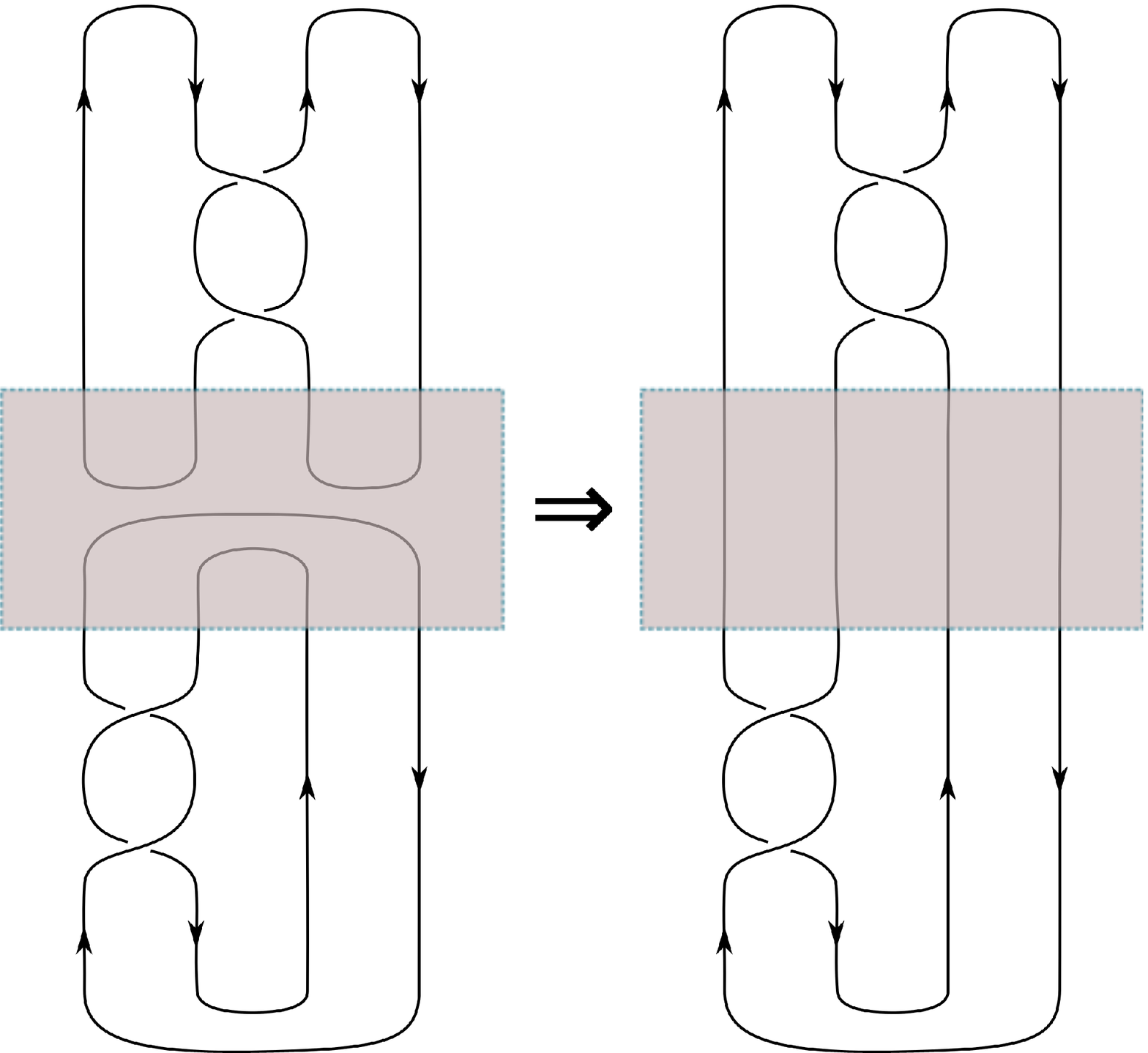}
\end{minipage}
}
 \caption{Construction of the figure-eight in the TQFT approach (left). Figure-eight knot obtained by a connect-sum of two Hopf
 links (right).}
 \label{fig_figure8}
\end{figure}

The formula for the Hopf link does involve the Racah coefficients. This is also true for any torus knot of the series $(2m+1,2)$
or torus links $(2m,2)$ with $m>0$ in the four-strand case. The simplest non-trivial example is the figure-eight knot $4_1$
(figure~\ref{fig_figure8}). The invariant of this knot can be computed as a double sum
\begin{multline}
\label{41Rmatrix}
H_R(4_1)=\sum\limits_{R_i,R_j\in R\otimes\bar{R}}\dim_q (R_i) q^{-2c_i}
\\ \times \left\{
\begin{array}{ccc}
R & \bar{R} & R_i \\
R & \bar{R} & R_j
\end{array}
\right\}\dim_q (R_j) q^{2c_j} \,,
\end{multline}
where the expression in curly brackets denotes the Racah matrix.

For example, in the fundamental representation of $SU(2)$ (\ref{41Rmatrix}) gives the following Jones polynomial
\be
J_{[1]}(4_1)=\frac{1+q^2}{q}\left(q^4-q^2+1-q^{-2}+q^{-4}\right)\,,
\ee
where we used the following expression for the Racah coefficients, \emph{cf.}~\cite{Zodinmawia:2011oya},
\be
\label{RacahFundamental}
\left\{
\begin{array}{ccc}
R & \bar{R} & R_i \\
R & \bar{R} & R_j
\end{array}
\right\}=\frac{1}{\dim_q R}\left(
\begin{array}{cc}
 1 & 1 \\
 1 & (1-{[N]^2})^{-1}
\end{array}
\right).
\ee
Here we use the following convention for the quantum version of a number
\be
[n]=\frac{q^{n}-q^{-n}}{q-q^{-1}}\,.
\ee

Equation~(\ref{41Rmatrix}) can be seen as a product of a Hopf link and its mirror image twisted by the Racah matrix, inserted in
between. Pictorially we would like to illustrate the deformation of the product of two Hopf links as an operation of cutting and
gluing together the links as shown in figure~\ref{fig_figure8}. The operation is valid if the representations of the Hopf links
coincide. An important difference between formulae~(\ref{212Rmatrix}) and~(\ref{41Rmatrix}) is in the multiplicity of summation,
single in the first example and double in the second. Now we would like to construct a similar generalization for the matrix
integral.

For the torus link $(m,n)$ colored with representations $R_1$, \ldots, $R_L$ of $U(N)$ the HOMFLY polynomial is given by the
following matrix (eigenvalue) integral~\cite{matrixmodel,Brini:2011wi} representation:
\begin{multline}
\label{IntTorusL}
H_{R_1,\ldots,R_L}({\L_{m,n}})=\frac{1}{Z_{m/L,n/L}}\int d u \,
\prod_{i=1}^Ne^{-u_i^2L^2/4\hat{g}_s} \\
\times \prod\limits_{i<j}^N4\sinh\left(\frac{u_i-u_j}{2m/L}\right)\sinh\left(\frac{u_i-u_j}{2n/L}\right)
\prod\limits_{i=1}^{L}\chs[R_i](e^u)\,.
\end{multline}
where
\be
\hat{g}_s=mn g_s, \qquad q=e^{g_s}\,,
\ee
$\chs[R](e^u)$ is the character of $U(N)$, which is given by the Schur polynomial and $Z_{m,n}$ is the normalization factor given
by the same integral with $\chs[R]\to 1$. The integral is taken over the eigenvalues $u_i$, $i=1,\ldots,N$ of the $U(N)$
matrices. The $U(1)$ part can be factorized out and we will be interested only in the $SU(N)$ part, what amounts to an
appropriate renormalization of characters.

In the following we use the ``correlator'' notation
\begin{multline}
\langle O(x)\rangle_{m,n,L} = \int d x \,
\prod_{i=1}^Ne^{-x_i^2L^2/4\hat{g}_s}
\\ \times \prod\limits_{i<j}^N4\sinh\left(\frac{x_i-x_j}{2m/L}\right)\sinh\left(\frac{x_i-x_j}{2n/L}\right) O(x)\,
\end{multline}
and
\be
\nav[O(x)]_{m,n,L}=\frac{\langle O(x)\rangle_{m,n,L}}{\langle 1\rangle_{m,n,L}}\,,
\ee
for the normalized version. In particular, for the Hopf link colored with the spin $k/2$ representation of $SU(2)$ one finds the
following Jones polynomial
\begin{multline}
\label{HlinkSU2}
J_{k,k}(2_1^2)=\nav[{\chs[k](e^{x})^2}]_{2,2,2}
\\ =\frac{q}{q^2-1}\int_{-\infty}^{\infty} d x \, e^{-x^2/8{g}_s} \left({e^{(k+1)x/2}-e^{-(k+1)x/2}}\right)^2
\\ =\sum\limits_{j=0}^{k^2+2k}q^{2j}\,.
\end{multline}

Going from the Hopf link to the figure-eight knot we propose the following generalization of the matrix integral formula, where
the product of two matrix integrals is twisted by some (non-local) kernel insertion:
\begin{multline}\label{MST}
\frac{1}{\langle\,1\,\rangle_{2,2,2}\langle\,1\,\rangle_{2,-2,2}}
 \int dx\,dy\,e^{-\frac{x^2}{4g_1}-\frac{y^2}{4g_2}}
\\ \times \left(\prod_{i<j}^N4\sinh\left(\frac{x_i-x_j}{2}\right) \sinh\left(\frac{y_i-y_j}{2}\right)\right)^2
\\ \times \chs[R](e^x)\chs[\bar{R}](e^x)\,G_R(x,y)\,\chs[R](e^y)\chs[\bar{R}](e^y)\,,
\end{multline}
where function $G_R(x,y)$ glues and twists two braids, representing $2_1^2$ links, in such a way that we get the $4_1$ knot. Here
we introduce two coupling constants $g_1$ and $g_2$ to distinguish the link and its mirror. After integration one sets
$e^{g_1}=q=e^{-g_2}$. If $G_R(x,y)=1$ we recover the result for the product of two links given by~(\ref{HlinkSU2}).

There is a simple, mirror-symmetric ($q\to 1/q$) basis for the colored Jones polynomials of the $4_1$ knot. This is given by
double matrix integrals of the form
\begin{multline}
\label{basis0}
\nav[S_k(e^{x})S_k(e^{y})]= \frac{1}{\langle\,1\,\rangle_{2,2,2}\langle\,1\,\rangle_{2,-2,2}}
\\ \times \int
dx\,dy\,\exp\left(-\frac{x^2}{4g_1}-\frac{y^2}{4g_2}\right) S_k(e^x)\,S_k(e^y)
\\ \times
\left(4\sinh\left(\frac{x_1-x_2}{2}\right) \sinh\left(\frac{y_1-y_2}{2}\right)\right)^2.
\end{multline}
Here the ``correlator'' of the two characters is normalized by two normalization factors of the $2_1^2$ Hopf link and  of its
mirror image. The integral yields
\begin{multline}
\label{correlatorbasis}
\nav[S_k(e^{x})S_k(e^{y})]=[k+1]^2 =q^{2k}+2q^{2k-2}+\ldots
\\  + kq^{2} +(k+1) + kq^{-2} + \ldots + 2q^{2-2k}+q^{-2k}\,.
\end{multline}

This basis can be related to the standard monomial basis using the Jones polynomial of $4_1$ in the representation $k/2$ as an
example. We have\footnote{With a slight abuse of notation we label $[p]$ the spin-$p/2$ representation of $SU(2)$. Otherwise
$[\cdot]$ stand for $q$-numbers.}
\begin{multline}
\label{41Jones}
\dim_q[p]\,J_{p}(4_1)=a_0+ \sum\limits_{i=1}^n a_i\left(q^{2i}+q^{-2i}\right)\,,
\\  n=p(p+2)\,.
\end{multline}
Then the correlator, which reproduces this result is
\begin{multline}
\label{41JonesCExpansion}
\nav[\sum\limits_{k=0}^n b_k S_k(e^x)S_k(e^y)]\,,
\\ \text{where} \qquad b_k=(A^{-1})_{ki}a_i\,,
\end{multline}
and $A$ is a non-degenerate $(n+1)\times(n+1)$ matrix
\be
A=\left(
\begin{array}{ccccc}
n+1 & n & n-1 & \cdots & 1\\
0 & n & n-1 & \cdots & 1 \\
0 & 0 & n-1 & \cdots & 1 \\
& & \cdots & & \\
0 & 0 & 0 & \cdots & 1
\end{array}
\right).
\ee

It is interesting to note the following property of the correlators, which immediately follows from the comparison of
(\ref{correlatorbasis}) and (\ref{HlinkSU2}):
\be
\label{identity}
\nav[S_k(e^x)^2S_k(e^y)^2]=\nav[S_{k(k+2)}(e^x)S_{k(k+2)}(e^{y})]\,.
\ee
The lhs of this expression is the invariant of the product of a link and its mirror image. The rhs is the highest order term in
expansion~(\ref{41JonesCExpansion}) of the Jones polynomial~(\ref{41Jones}). The above property shows us how the invariant of the
knot~(\ref{41Jones}) can be understood as a deformation of the formula for the product of two links. One can also notice that the
number of terms $(p+1)^2$ in the sum~(\ref{41Jones}) coincides with the number of
terms in the sum~(\ref{41Rmatrix}).

Thus, the Jones polynomials can be found as an expansion
\be
\label{matrixrep}
\dim_q[p]J_{[p]}(4_1)= \sum\limits_{k=0}^{p(p+2)}b_k(p)[k+1]^2\,, \qquad
\ee
with some integer coefficients $b_k$, which depend on the representation $[p]$:
\begin{multline}
b_{p^2+2p}=1\,, \qquad b_{p^2+2p-1}=-1\,,
\\  b_{p^2+2p-2}=-1\,, \qquad \ldots
\end{multline}

Since the general formula for colored Jones polynomials of the figure-eight can be found in~\cite{Itoyama:2012fq}, this allows
us to find the coefficients $b_k$ in any representation of $SU(2)$. In particular, one can derive some recursive relations for
the coefficients~\cite{OurFutureWork}.

As follows from~(\ref{41Rmatrix}) the coefficients $b_k$ satisfy the resummation formula involving the Racah coefficients
\begin{multline}
\label{RacahResummed}
\dim_q[p]J_{[p]}(4_1)= \sum\limits_{k=0}^{p(p+2)}b_k[k+1]^2
\\ =\sum\limits_{i,j=0}^{p}[2i+1][2j+1]q^{2i(i+1)-2j(j+1)}\alpha_{ij}(p)\,,
\end{multline}
where
\be
\alpha_{ij}(p)=\dim_q[p]\left\{
\begin{array}{ccc}
 {[p]} & [p] & {[2i]}\\
 {[p]} & [p] & {[2j]}
\end{array}
\right\}.
\ee

\begin{figure}
 \centering
 \includegraphics[width=0.9\linewidth]{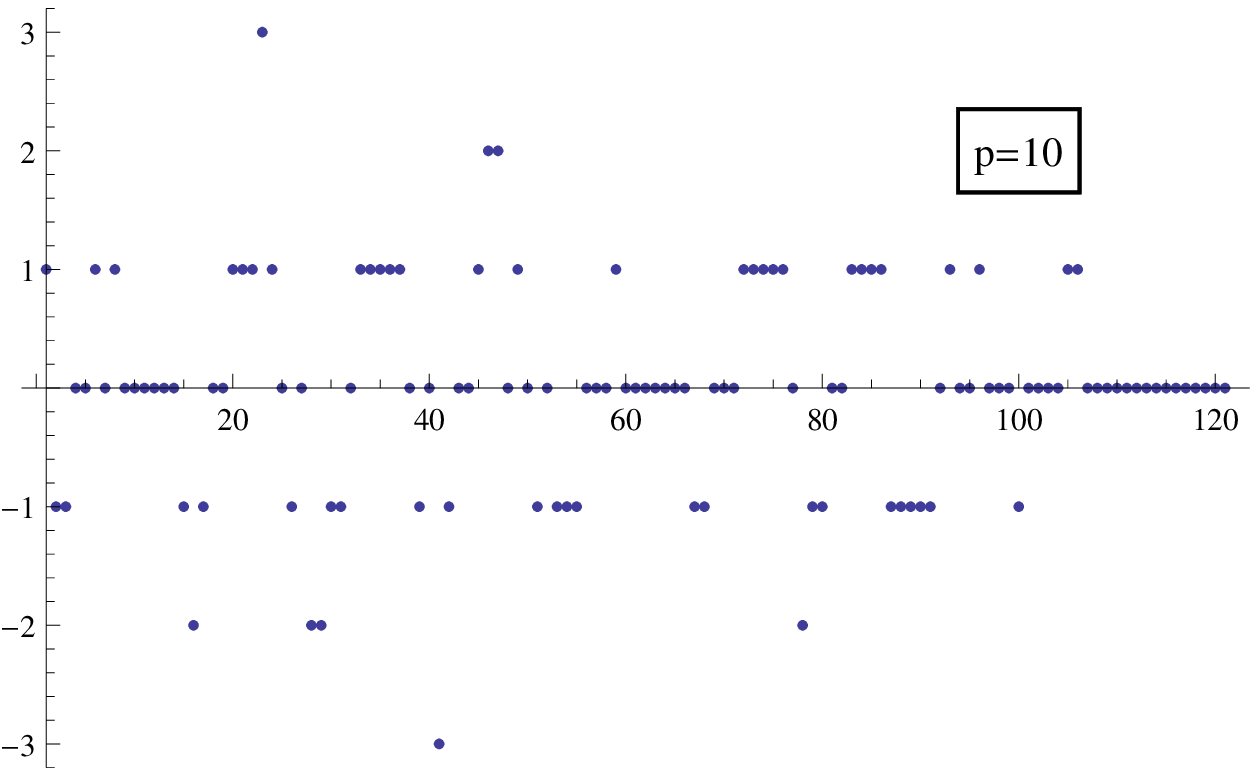}
 \vspace{0.5cm}

 \includegraphics[width=0.9\linewidth]{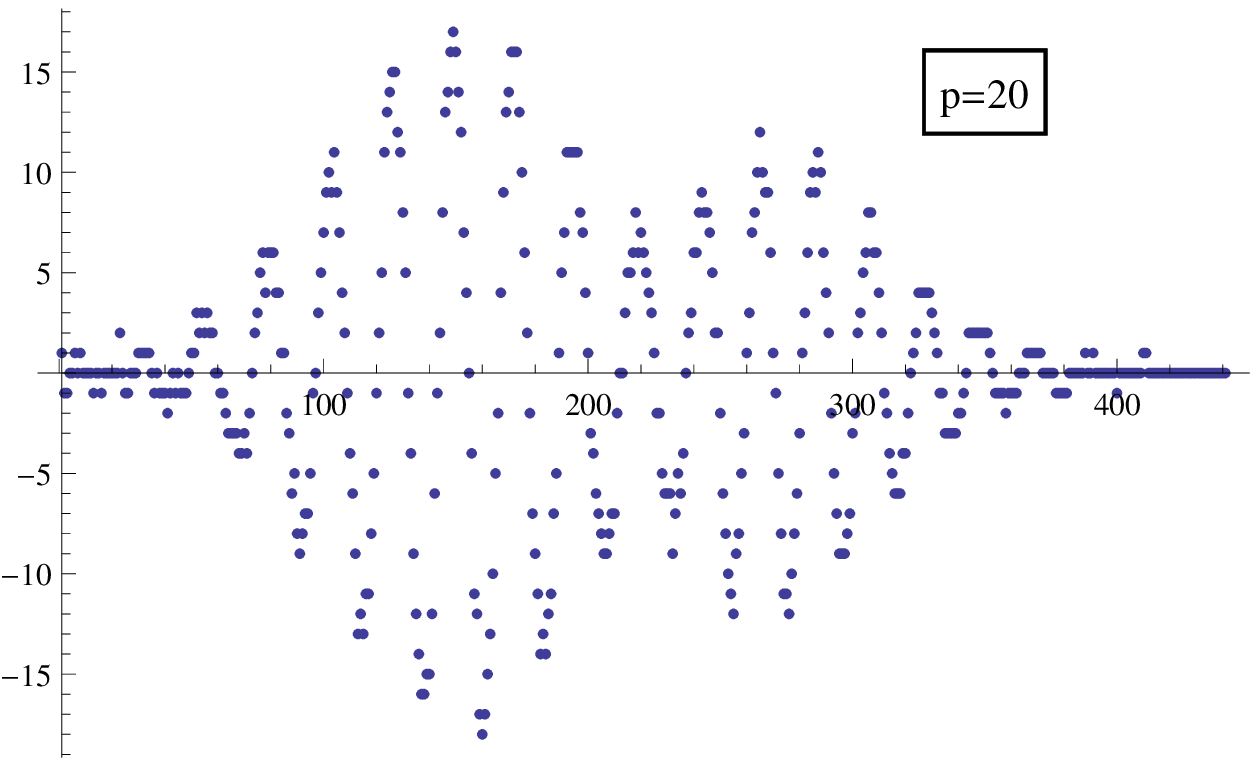}
 \vspace{0.5cm}

 \includegraphics[width=0.9\linewidth]{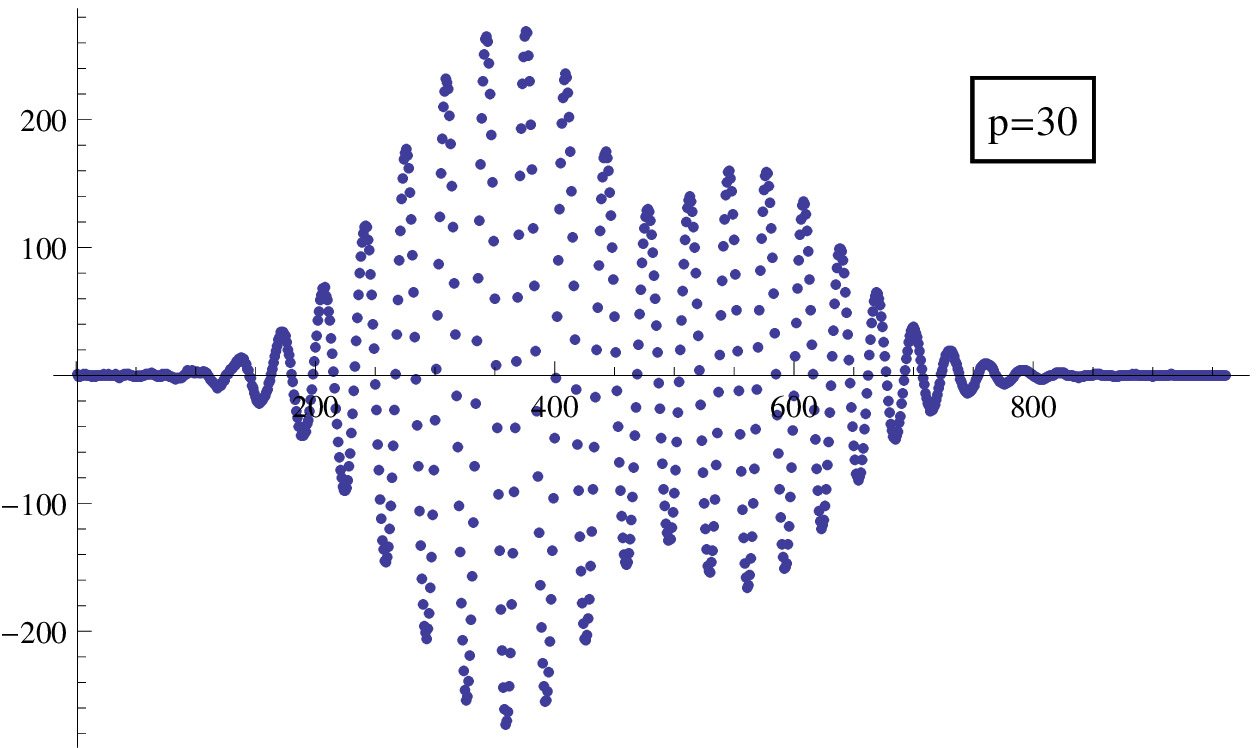}
 \vspace{0.5cm}

 \includegraphics[width=0.9\linewidth]{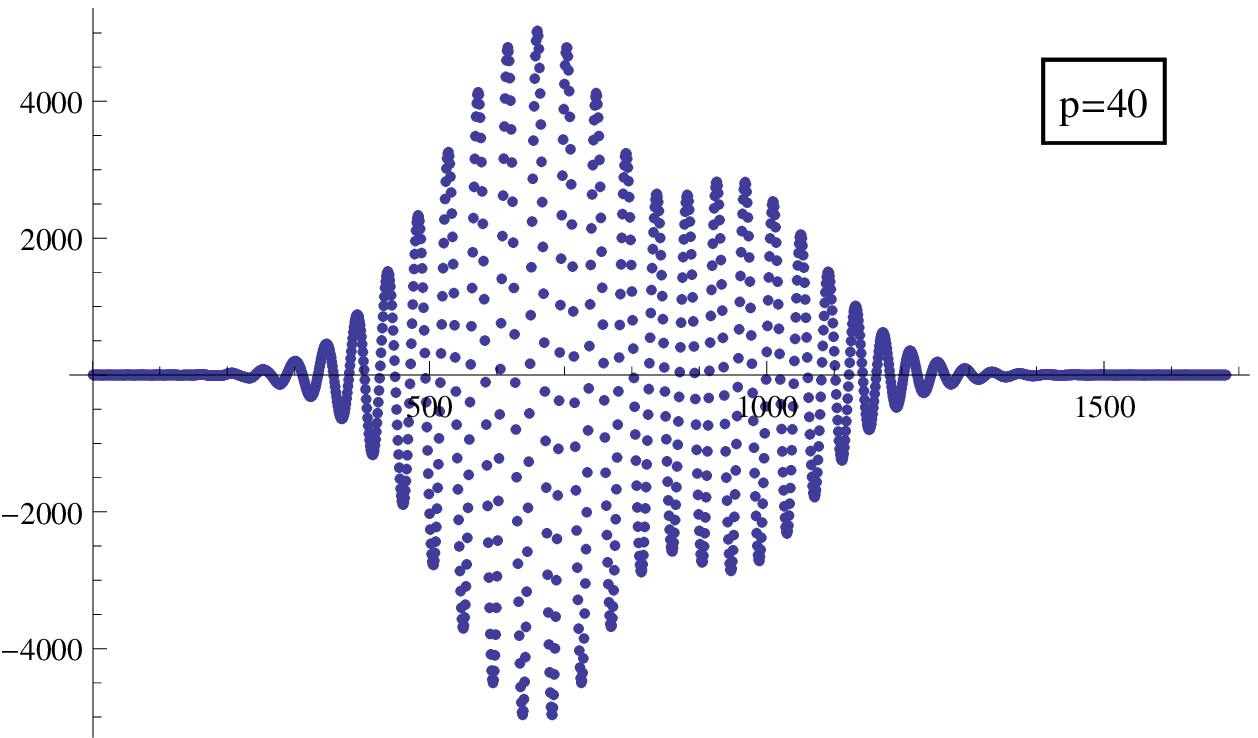}

 \caption{Value of the coefficients $b_k$ as a function of $k$ in representations $[10]$, $[20]$, $[30]$ and $[40]$}
 \label{fig_distrib}
\end{figure}

Figure~\ref{fig_distrib} shows the distribution of the coefficients in the expansion of $\dim_q[p]J_p(4_1)$ for $p=10,20,30,40$.
The plots show an interesting pattern for the distributions for higher representations. They also hint at another interesting
property of the coefficients $b_k(p)$,
\be
\sum b_k(p) = \left\{ \begin{array}{l}
                       0, \quad \text{if $p$ is odd} \\
                       1, \quad \text{if $p$ is even}
                      \end{array}

\right..
\ee

The generalization of formula~(\ref{41Rmatrix}) for the $(2m)_1$ knot series reads
\begin{multline}
\label{2m1Rmatrix}
H_R(2m_1)=\sum\limits_{R_i,R_j\in R\otimes\bar{R}}\dim_q (R_i) q^{-2c_i}
\\ \times \left\{
\begin{array}{ccc}
 R & \bar{R} & R_i \\
 R & \bar{R} & R_j
\end{array}
\right\}\dim_q (R_j) q^{(2m-2)c_j} \,.
\end{multline}
Apparently, in the general case the ``mirror symmetry" is broken and the invariants will be no longer symmetric with respect to
$q\to 1/q$.

One can also try to present the invariants of the knot from the $2m_1$ series as a deformation of the product of 2 links, a
(mirror image of a) Hopf link and a $(2m-2)_1^2$ link. Indeed the  generalized correlator $\nav[S_k(e^{x})^2S_k(e^{y})^2]_m$
gives the leading order contribution to $J_{k}(2m_1)$. However, the property~(\ref{identity}) does not hold for
$\nav[S_k(e^{x})S_k(e^{y})]_m$. Correlators $\nav[S_k(e^{x})S_k(e^{y})]_m$ do not form a basis for the invariants for general
$m$. It would be interesting to find the appropriate generalization of the $m=1$ basis, which will be done
elsewhere~\cite{OurFutureWork}.

The authors would like to thank Andrei Mironov and Alexei Morozov for the collaboration at the initial stage of this project and
for the numerous insightful conversations. The work of AA was supported in part by RFBR grant 14-01-00547, NSh-1500.2014.2 and by
the Federal Agency for Science and Innovations of Russian Federation.  The work of DM was partly supported by the Brazilian
Ministry of Science, Technology and Innovation (MCTI) and by Science without Borders program of the CNPq, The National Council of
the Scientific and Technological Development-Brazil, the MIT-IIP exchange program, by the Russian RFBR grant 14-02-00627,
binational cooperation grant RFBR-India 14-01-92691\_Ind\_a and the grant for support of Scientific Schools NSh 1500.2014.2. DM
would also like to thank the hospitality of the Freiburg Institute of Advanced Studies, where this work was initiated.

\end{document}